\documentclass[doublecol]{epl2}
\usepackage{amsmath}
\usepackage{graphicx}
\usepackage{epsf}
\usepackage{psfrag}
\usepackage{epsfig}
\usepackage{graphics}

\newcommand{\be}{\begin{equation}}
\newcommand{\ee}{\end{equation}}
\newcommand{\bq}{\begin{eqnarray}}
\newcommand{\eq}{\end{eqnarray}}

\newcommand{\dena}{(k^2-\mu^2)}
\newcommand{\intk}{\int_{\vec{k}}}

\title{Conductivity of Coulomb interacting massless Dirac particles in graphene: regularization dependent parameters and symmetry constraints}
\shorttitle{Conductivity of Coulomb interacting massless Dirac particles in graphene}
\author{G. Gazzola\inst{1} \and A. L. Cherchiglia\inst{1}\thanks{E-mail: \email{adriano@fisica.ufmg.br}} \and L. A. Cabral\inst{2} \and
M. C. Nemes\inst{1} \and Marcos Sampaio\inst{1}}
\institute{
\inst{1} Universidade Federal de Minas Gerais - Departamento de F\'{i}sica - ICEx 
P.O. BOX 702, 30.161-970, Belo Horizonte MG - Brazil\\
\inst{2} Universidade Federal do Tocantins - Departamento de F\'{i}sica 
P.O. Box 132, 77804-970, Aragua\'{\i}na, TO - Brazil
}
\shortauthor{G. Gazzola \etal}
\pacs{72.80.Vp}{Graphene, eletronic transport}
\pacs{11.10.Gh}{Renormalization and regularization}

\abstract{We compute the Coulomb correction $\mathcal{C}$  to the a. c. conductivity of interacting massless Dirac particles in graphene in the collisionless limit using the polarization tensor approach in a regularization independent framework. Arbitrary parameters stemming from differences between logarithmically divergent integrals are fixed on physical grounds exploiting only spatial $O(2)$ rotational invariance  of the model which amounts to transversality of the polarization tensor. Consequently $\mathcal{C}$ is unequivocally determined to be $(19- 6\pi)/12$ within this effective model. We compare our result with explicit regularizations and discuss the origin of others results for $\mathcal{C}$ found in the literature.}

\begin{document}

\maketitle

\section{Introduction}
\label{Introducao}
The value of graphene a.c. conductivity corrected by Coulomb interaction  has been a matter of debate in recent literature. Its general form for low frequencies can be obtained in the context of renormalization group techniques \cite{Herbut1},\cite{Sheehy2},\cite{Mishchenko2} based on scale relations valid near the quantum critical point:
\begin{equation}
	\sigma(\omega) = \sigma_0 \left( 1 + {\cal C}\frac{e^2}{v_F + \frac{e^2}{4}\ln{\frac{\Lambda}{\omega}}} \right) .
\label{condutividade}
\end{equation}
In the equation above $\sigma_0=e^2/4\hbar$ is known as ``minimum conductivity'' (in absence of interactions), measured in \cite{Kuzmenko1}, \cite{Science} as $\sigma_0=(1,01\pm0,04)e^2/4\hbar$, $v_F$ is the Dirac fermion velocity, which is renormalized by $\frac{e^2}{4}\ln{\frac{\Lambda}{\omega}}$ due to the Coulomb interaction and $\Lambda$ is an upper cutoff \cite{Vozmediano1}. 
This is the case of the  coefficient ${\cal C}$  which can be calculated within a model that takes Coulomb interaction into account. In order to obtain ${\cal C}$,  the majority of calculations rely on a perturbative analysis of effective theories based on disorder-free Coulomb interacting massless Dirac electrons. However a complete eletronic structure calculation based on a realistic tight-binding Hamiltonian has  been performed in \cite{Rosenstein}. In diagrammatic perturbation theory to first order in electron-electron interaction, Feynman diagrams that contribute to the density-density and current-current response function (expressed together by a polarization tensor) can be drawn in analogy to field theoretical quantum electrodynamics (QED) \cite{Vignale}. The general structure of the polarization tensor can be established on geometric grounds as well as a continuity equation $ \nabla\cdot\vec{j}+\frac{\partial\rho}{\partial t}=0 $ which  expresses the density-density response function in terms of current-current response function. Although the geometry of the graphene honeycomb lattice is $C_6$ symmetric, the low effective model to be treated here is spatially $O(2)$ symmetric. Thus while rotational invariance leads to transversality of the polarization tensor, the continuity equation  constrains the form of the Coulomb vertex function leading to a Ward-Takahashi like identity similar to QED, namely  a relation between the vertex function and the electron self-energy \cite{Juricic1}, \cite{Sheehy1}. It was claimed, however, that while the continuity equation holds at noninteracting level, it fails when electron-electron interactions are taken into account \cite{Vignale} because of  ultraviolet infinities which demands a rigid momentum cutoff. 

The main results available in the literature within effective models are obtained using the Kubo Formula \cite{Herbut1}, \cite{Juricic1}, \cite{Sheehy1} \cite{Mishchenko1}, Electron Polarization Operator \cite{Juricic1}, \cite{Mishchenko1} and Kinetic Equation \cite{Mishchenko2},\cite{Mishchenko1} yielding different results for ${\cal C}$ depending on how intermediate divergent integrals are handled in each method.

%

The ab initio tight binding method of \cite{Rosenstein} claims that the ambiguity characterizing the various approaches is related to a chiral anomaly in the system in consequence of an unclear separation between infrared and ultraviolet physics and thus the regularization of an effective model is a crucial and important issue. The authors in \cite{Rosenstein} also point out that their result is compatible with \cite{Juricic1} namely ${\cal{C}}=\frac{11-3\pi}{6}$ where dimensional regularization was employed even though it is known to be problematic in $2+1$ anomalous theories and needs to be handled with care when describing chiral anomalies \cite{Bertlmann}, \cite{Jackiw1}. It is well known that in the presence of anomalies such as the Adler-Bardeen-Bell-Jackiw (ABJ) chiral anomaly \cite{ABBJ} a regularization should play no role in picking a particular axial or vector Ward identity but rather should display the anomaly in a manifestly ``democratic way" to distinguish spurious from physical anomalies.  

 In \cite{Juricic1} the authors used the polarization operator approach and dimensional regularization to handle intermediate divergent integrals in the calculation of the conductivity. They claimed that their previous result presented in \cite{Herbut1} lacked a consistent dimensional continuation of Pauli matrices and thus violated a  Ward identity based on the continuity equation. Moreover a missing term (absent in the physical dimension) related to space-time continuation of the sigma matrix algebra was argued to account for the discrepancy between the results for ${\cal C}$, $\frac{25-6\pi}{12}$ in \cite{Herbut1} and $\frac{11-3\pi}{6}$ in \cite{Juricic1}.

The purpose of the present contribution is to shed some light on this controversy employing the polarization operator method. Our strategy is based on a regularization independent approach namely implicit regularization (IR) that operates in the physical dimension of the model.  An important feature of IR which makes it an ideal arena to calculate the conductivity is that it clearly parametrizes finite regularization dependent terms which in our approach stem from differences between logarithmically divergent integrals (surface terms - ST) containing one integration variable only (loop momentum). This appears to be at the heart of controversial results for ${\cal C}$.
Such arbitrary regularization dependent ST should be fixed on physical grounds employing symmetry requirements of the underlying model or phenomenology. This is especially relevant when we work with radiative corrections that are finite although intermediate divergent integrals appear because the original amplitude is superficially divergent. This is the case of the effective models used to compute the conductivity. 

As we shall demonstrate in this paper, we obtain for the a.c. conductivity using the current-current correlator,
\begin{equation}
	{\cal C} = 2\pi\alpha+\frac{19-6\pi}{12},
\label{resultado C reg implicita}
\end{equation}
where $\alpha$ is an arbitrary finite constant (ST) which parametrizes the regularization dependence and whose expression we derive in the next sections. In the present case we will demonstrate that $\alpha$ is unambiguously fixed to zero {\cal{only}} by the general transversal structure of the polarization tensor inferred from spatial $O(2)$ rotational invariance. Consequently we have ${\cal C} = \frac{19-6\pi}{12}$. Two features are noteworthy. Firstly the continuity equation (whose validity is controversial when interactions are considered \cite{Vignale}) which leads to Ward Takahashi like identity for the vertex function in \cite{Juricic1} plays no r\^ole in determining $\alpha$. Secondly  $\alpha$ evaluates to zero in dimensional regularization which means that in principle such regularization agrees with our result as well yet this is not the value obtained in \cite{Juricic1}. From the experimental point of view the best testground for ${\cal{C}}$, whose controversial values differ by one order of magnitude, resides in optical transparency of graphene as the transmittance is related to the conductivity in the optical regime as $t(\omega) = (1+2 \pi \sigma(\omega)/c)^{-2}$. Experimental data on the optical transmission \cite{Science} suggest a negligible correction due to Coulomb interaction compatible with the value ${\cal C} = \frac{19-6\pi}{12} \approx 0.01$. Furthermore, according to \cite{Science} the optical properties of graphene lie mainly on its two dimensional structure and gapless eletronic spectrum and does not involve the chirality of the charge carriers \cite{Rosenstein}. 
Moreover the model remains predictive because transversality derives from a very general symmetry property of the model which remains valid in the presence of interactions. We also map the different values of ${\cal{C}}$ to different evaluations of the regularization dependent parameter $\alpha$ which is per se arbitrary and should be fixed on symmetry requirements. 


\section{The formalism: electrical conductivity from imaginary time correlation function}
\label{AC Conductivity Revisited}

 The Hamiltonian of monolayer graphene is described by two dimensional massless Dirac (quasi)particles with the speed of light replaced by $v_F$ and the pseudo-spin corresponding to sublattice indices \cite{Neto1},
 \bq
 {\hat{H}} &=& \int d^2\vec{r} \, \psi^\dagger(\vec{r})v_F \vec{\sigma}\cdot\vec{p}\psi(\vec{r}) \nonumber\\ &+& e^2\int d^2\vec{r}d^2\vec{r'}
 \frac{\psi^\dagger(\vec{r})\psi(\vec{r}) \psi^\dagger(\vec{r'})\psi(\vec{r'})}{|\vec{r}-\vec{r'}|},
\label{Hamiltoniana}
 \eq
 where $\vec{\sigma}=(\sigma_x,\sigma_y)$ are Pauli matrices and we included a two-body Coulomb interaction between electrons.

 Quantum electrodynamics in 2+1 dimensions can be used to describe planar fermions in two spatial dimensions to calculate, for instance, the polarization operator defined through the effective action of fermions in the presence of electromagnetic fields. The polarization tensor can be interpreted in terms of the conductivity of graphene \cite{Natures} as well as for the description of other interesting phenomena such as the Hall and Faraday effects, the light absorption rate by graphene sheets and the Casimir interaction of graphene \cite{Fialkovsky2}, \cite{Vozmediano3}.  

Based on the fluctuation dissipation theorem \cite{KUBOM}, the expectation value of the electrical current density operator for real time  $j(\vec{r}, t)$ can be related to the imaginary time  polarization tensor $\Pi_{\mu\nu}(\tau, \vec{r}) = \langle T [j_\mu (\tau, \vec{r})j_\nu(0,0)]\rangle$ with $j_\mu (\tau, \vec{r}) = (\psi^\dagger (\tau, \vec{r}) \psi (\tau, \vec{r}), v_F \psi^\dagger (\tau, \vec{r}) \vec{\sigma} \psi (\tau, \vec{r}))$,
$\mu = 0, 1, 2$ and $\psi (\tau, \vec{r})$ is the two component massless field.

The correlation function in the reciprocal space $\Pi_{\mu\nu}(q_\mu)$, $q_\mu = (i \Omega, \vec{q})$ and $\vec{q}=(q_1,q_2)$, will be calculated within an expansion in the coupling constant to order $e^2$ just as in \cite{Juricic1}, namely
\begin{equation}
	\Pi_{\mu\nu}(q_\mu)=\Pi^0_{\mu\nu}(q_\mu)+\delta\Pi_{\mu\nu}(q_\mu,V_{\vec{k}}),
\label{Pimunu expansao}
\end{equation}
where $\Pi^0_{\mu\nu}$ represents the noninteracting contribution which characterizes the minimum conductivity, $\delta\Pi_{\mu\nu}$ is the first correction due to  Coulomb interaction of the quasiparticles and $V_{\vec{k}}=\frac{2\pi e^2}{|\vec{k}|}$ is the Fourier transform of the Coulomb potential.
The leading order term in \eqref{Pimunu expansao} reads
\begin{equation}
	\Pi^{0}_{\mu\nu}(q_\mu)=-4\int_\omega \int_{\vec{k}}{\textnormal{Tr}}[G_{\omega, k}\sigma_\mu G_{\omega+\Omega, k+q}\sigma_\nu]
\label{Pi0munu definicao}
\end{equation}
in which $\int_\omega \equiv \int_{-\infty}^{+\infty} d\omega/(2\pi)$, $\intk \equiv \int d^2k/(2 \pi)^2$, $\sigma_\mu = ( 1_{2 \times 2}, \sigma_x, \sigma_y)$, $4$ is the number of two component fermionic fields copies for the graphene and $G_k=\frac{i\omega+\vec{\sigma}\cdot\vec{k}}{\omega^2+\vec{k}^2}$ is the fermion propagator
In turn the correction term to the correlation function can be written as
\be
	\delta\Pi_{\mu\nu}(q_\mu) = 4 \int_{\vec{k},\vec{p},\omega,\omega'}  V_{\vec{k}-\vec{p}} {\textnormal{Tr}}[\mathcal{A}_1
+ \mathcal{A}_2 + \mathcal{A}_3 ],
\label{correcao}
\ee
in which
\bq
\mathcal{A}_1 &\equiv& G_{\omega, k}\sigma_\mu G_{\omega+\Omega, k+q}G_{\omega'+ \Omega, p+q}\sigma_\nu G_{\omega', p},\nonumber\\
\mathcal{A}_2 &\equiv& G_{\omega, k}\sigma_\mu G_{\omega +\Omega, k+q}\sigma_\nu G_{\omega, k}G_{\omega', p}\,\,\, \mbox{and} \nonumber\\
\mathcal{A}_3 &\equiv& G_{\omega, k}\sigma_\mu G_{\omega +\Omega, k+q}G_{\omega'+\Omega, p+q}G_{\omega+\Omega, k+q}\sigma_\nu .\nonumber
\eq
The correction to the conductivity may be derived taking the spatial component of the polarization tensor (current-current approach)
\begin{equation}
	\sigma(\Omega,|q|)=\frac{e^2}{\hbar}\frac{i\Omega}{q^2-\Omega^2}\Pi_{xx}(\Omega+i0,|q|),
\label{Relacao Pixx e sigma}
\end{equation}
the a.c. conductivity being given by $\lim_{q\to0}\sigma(\Omega,|q|)$ \cite{Juricic1,Mishchenko1}.

\section{Implicit regularization and the parametrization of (un)determined  perturbative corrections}
\label{Regularizacao Implicita}

Implicit regularization (IR) is a momentum space framework that operates on the physical space-time dimension of the underlying model. It assumes an implicit regularization operating in the physical space-time dimension for a general $n$-loop Feynman amplitude only to allow the application of a mathematical identity at the level of propagators which displays its divergent content as basic divergent integrals (BDI)'s that are written in terms of one internal momentum only. 
We suggest references \cite{AnomalyRI}, \cite{Ireg} for a complete account of IR and discuss here some basic features related to leading order calculations in two dimensions.

For instance in the case where the ultraviolet behavior is logarithmical, an one loop Feynman amplitude is cast as a finite function of external momenta, a BDI (say $I_{log} (\lambda^2)$) and surface terms expressed by integrals of a total derivative in momentum space. The origin of these surface terms is that logarithmically divergent loop integrals $I_{log}^{\mu \nu \ldots} (\lambda^2)$ which contain in the integrand a product of internal momenta carrying Lorentz indices $\mu, \nu, \ldots $ can be expressed in a precise way  as a product of metric tensors symmetrized in the Lorentz indices and $I_{log} (\lambda^2)$ plus a surface term. Such local, regularization dependent surface terms are intrinsically arbitrarily valued. 
Within IR, regularization dependent terms (surface terms) can be extracted out in a consistent way allowing for a clear discussion of the ambiguities involved in the manipulation of divergent integrals. Because it acts  on the physical dimension of the theory, IR is particularly useful to dimensional specific models. In the latter, dimensional regularization methods flaw  because of ambiguities in the analytic continuation on the space-time dimension. 

To illustrate this method in connection with our calculation of the a.c. conductivity in graphene, we discuss the quantum mass generation for photons in quantum electrodynamics in 1+1 Minkowski space-time dimension (Schwinger Model). We consider massless fermions for simplicity. This is a good example as  in the calculation of graphene a.c. conductivity, after integrating in the frequencies, we end up with  momentum integrals in 2-dimensional Euclidean space. Moreover, the photon mass generation is evaluated through the calculation of the vacuum polarization tensor 
\be \Pi_{S}^{\mu \nu}
(p) = i \mbox{Tr} \int_k \, \gamma^\mu
\frac{i}{k\hspace{-2mm}/}\gamma^\nu
\frac{i}{k\hspace{-2mm}/+p\hspace{-2mm}/} \, . \label{pmnir}
\ee
\noindent
which turns out to be finite even though it is superficially (logarithmically) divergent. This happens to be the case in the polarization tensor and vertex function for the calculation of the conductivity as well. 


We evaluate this amplitude using IR. After performing the trace algebra, we separate the divergent content in terms of the loop momentum using the identity
\be
\frac{1}{[(k+p)^2-\mu^2]} = \frac{1}{(k^2-\mu^2)}\Bigg[1 - \frac{(p^2 + 2 p\cdot k)}{(k+p)^2-\mu^2}\Bigg] \, ,
\label{identidade fundamental}
\ee
$\mu$ being a fictitious mass for the electron which will be taken to zero in the end. Then it is easy to show that \cite{AnomalyRI}
\be \Pi_{S}^{\mu \nu} (p) =
\frac{1}{\pi} \Big(\frac{\alpha + 2}{2}g^{\mu \nu} - \frac{p^\mu
p^\nu}{p^2}\Big) \, , \label{pmnsm} \ee
in which the arbitrary {\it{regularization dependent}} parameter $\alpha$ is the difference between two logarithmically divergent integrals
\bq
\alpha g_{\mu \nu}  &\equiv&
g_{\mu \nu} I_{log}(\mu^2) - 2 I_{log \, \mu \nu} (\mu^2) \nonumber \\
&=& \int_k \frac{g_{\mu \nu}}{\dena} -
2 \int_k \frac{k_\mu k_\nu}{\dena^2} \nonumber \\
&=&
\int_k \frac{\partial}{\partial k^\nu}\Big(\frac{k_\mu}{(k^2-\mu^2)}\Big)
\equiv
\Xi_{\mu \nu}.
\label{RC}
\eq
If explicitly calculated in dimensional regularization or Pauli-Villars it evaluates to zero in consonance with the transversal character of the polarization tensor required by gauge invariance, whereas $\alpha = 1/(4 \pi)$ in sharp-cutoff. The resulting radiatively generated photon mass is $m_\gamma^2 = e^2/\pi$. However, following Jackiw in \cite{Jackiw1} we can consider $\alpha$ as an arbitrary undetermined parameter. This is a good example of radiative corrections that are finite. As such they should be fixed on symmetry or phenomenological grounds. Thus, in this case, we can assign a vanishing value for $\alpha$ based either on transversality or the ``phenomenological" value of the generated photon mass should it exist at all. However, in its chiral version, the Chiral Schwinger model exhibits an anomalous non-simultaneous conservation of the vector and chiral current \cite{Shifman1} similar to the famous ABJ triangle anomaly \cite{ABBJ}. The democratic display of the anomaly between the two Ward identities can only be achieved if $\alpha$ is left arbitrary. The calculation by itself should not decide which Ward identity is to be satisfied.

An arbitrary positive (renormalization group) mass scale $\lambda$ appears via a regularization independent identity which enables us to write a BDI as a function of $\lambda$ only plus logarithmic functions of $\mu/\lambda$. At one loop order it reads
\begin{equation}
 I_{log} (\mu^2) = I_{log} (\lambda^2) + \frac{i}{4 \pi} \ln \Big( \frac{\mu^2}{\lambda^2}\Big)\, .
\label{Relacao de Escala Ilog Minkowski}
\end{equation}

 For the sake of comparison with other regularizations, we can also construct an explicit general parametrization for basic divergent integrals  exhibiting explicitly all arbitrary regularization dependent parameters and revealing the divergent behavior through a cutoff. 
In order to construct such parametrization in $1+1$ space-time dimensions, consider the regularization independent derivatives with respect to $\mu^2$,
\be
\frac{d I_{log}(\mu^2)}{d \mu^{2}}= \frac{i}{4 \pi \mu^{2}}\quad,\frac{d I_{log}^{\mu \nu}(\mu^2)}{d \mu^{2}}= \frac{g^{\mu \nu}}{2}\frac{i}{4 \pi \mu^{2}}.
\ee

A general parametrization which obeys the relations above is given by
\begin{align}
I_{log} (\mu^2)&=  -\frac{i}{4 \pi} \ln\left(\frac{\Lambda^2}{\mu^2}\right) + c,\nonumber\\
I_{log}^{\mu\nu}(\mu^2)&= \frac{g^{\mu\nu}}{2}\Bigg[ - \frac{i}{4 \pi}\ln\left(\frac{\Lambda^2}{\mu^2}\right) + c'\Bigg],
\label{resilog}
\end{align}
\noindent
in which $c$, $c'$ are also arbitrary dimensionless regularization dependent constants and $\Lambda \rightarrow \infty$. The arbitrariness of the surface term defined in (\ref{RC}) becomes evident since $\Xi^{\mu \nu} = \alpha g^{\mu \nu} = (c-c')g^{\mu \nu}$. 

\section{Calculation of A.C. Conductivity}

Hereafter we shall work in Euclidian space. The minimum conductivity can be calculated from eq. (\ref{Pi0munu definicao}) after subtracting the zero frequence mode, namely $\Pi^{0 ,\Omega}_{\mu\nu}(q_\mu)\rightarrow \Pi^{0 ,\Omega=0}_{\mu\nu}(q_\mu)-\Pi^{0, \Omega}_{\mu\nu}(q_\mu)$ which eliminates a spurious linear divergence $I_{lin} (\mu^2)=\int_{\vec{k}}(k^2+\mu^2)^{-1/2}$. After some algebra we obtain
\bq
&&	{\Pi}^{0}_{\mu \nu}(q_\mu)=-\frac{\sqrt{\Omega^2+q^2}}{16} {\mbox{Tr}} \Big[ \sigma_\mu \sigma_\nu- 2\delta_{\mu 0}\sigma_\nu  \nonumber \\
 && - \frac{(i\Omega+\vec{\sigma}\cdot\vec{q})\sigma_\mu(i\Omega+\vec{\sigma}\cdot\vec{q})\sigma_\nu}{\Omega^2+q^2} \Big] + \frac{\sqrt{q^2}}{16} \times \nonumber \\&&  {\mbox{Tr}}\Big[ \sigma_\mu\sigma_\nu -4\delta_{\mu 0}\sigma_\nu - \frac{\vec{\sigma}\cdot\vec{q}\,\sigma_\mu \, \vec{\sigma}\cdot\vec{q} \, \sigma_\nu}{q^2} \Big]
\label{Pimunu resultado sem beta}
\eq
which gives from (\ref{Relacao Pixx e sigma}) the well known universal value of the conductivity $\sigma_0 = \frac{1}{4} \frac{e^2}{\hbar}$. The Coulomb correction to the conductivity can be calculated from eq. (\ref{correcao}), which we separate into two parts namely the self energy correction $\delta \Pi^a_{\mu \nu}$ and the vertex correction  $\delta \Pi^b_{\mu \nu}$,
\begin{equation}
\delta\Pi_{\mu\nu}(i\Omega,0)=\delta\Pi^{a}_{\mu\nu}(i\Omega,0)+\delta\Pi^{b}_{\mu\nu}(i\Omega,0)
\label{deltaPicmunu separacao}
\end{equation}
where we have already taken the limit $q \rightarrow 0$. Explicitly we have
\bq
	\delta\Pi^{a}_{\mu\nu}&=&8 \int_{\omega,\omega',\vec{k},\vec{p}} V_{\vec{k}-\vec{p}} \mbox{Tr}
\Big[G_{\omega, k}\sigma_\mu G_{\omega+\Omega, k}\sigma_\nu  \nonumber\\ &\times&
G_{\omega, k} G_{\omega', p} \Big] \,\,\,\ \mbox{and}
\label{deltaPiamunu definicao}
\eq
\bq
\delta\Pi^{b}_{\mu\nu}&=& 4 \int_{\omega,\omega',\vec{k},\vec{p}} V_{\vec{k}-\vec{p}} \mbox{Tr}
\Big[G_{\omega, k}\sigma_\mu G_{\omega+\Omega, k}\nonumber \\  &\times& G_{\omega'+\Omega, p}
\sigma_\nu  G_{\omega', p} \Big].
\eq
The integrals above can be simplified if we already take  $\mu = \nu = x$, the diagonal spatial component of the Coulomb interaction correction $\delta \Pi_{xx} (i\Omega, 0)$, to calculate the conductivity within the current-current approach. After subtracting the zero mode just as we did for the noninteracting case, we obtain, after tedious yet straightforward calculation, that $\delta\Pi^{a}_{xx}$ and $\delta\Pi^{b}_{xx}$ contribute to the conductivity with
\bq
	\frac{\sigma_a}{\sigma_0 e^2} &=& -\Big(\pi I_{log}(\lambda^2)+\frac{1}{4}\ln{\lambda^2}-\pi\alpha \nonumber \\ &+& \frac{3}{2}\ln{2}-\frac{1}{4}-\frac{1}{2}\ln{\Omega}\Big)
\label{sigmaa resultado}
\eq
and
\bq
&&	\frac{\sigma_b}{\sigma_0 e^2} = \Big(\pi I_{log}(\lambda^2)+\frac{1}{4}\ln{\lambda^2}+\pi\alpha + \frac{3}{2}\ln{2}\nonumber \\ &&- \frac{1}{2}\ln{\Omega} - \frac{\pi}{2} +\frac{1}{12}(4+3\pi) +\frac{4-\pi}{4} \Big), \label{sigmab resultado}
\eq
respectively. Thus the  correction to the conductivity is given by adding (\ref{sigmaa resultado}) to (\ref{sigmab resultado}) to yield
\begin{equation}
	{\cal{C}} = \left(2\pi\alpha+\frac{19-6\pi}{12} \right).
\label{sigmac resultado}
\end{equation}
Some comments are in order. Firstly some remarkable (regularization independent) cancellations take place.
Note that the logarithmic divergences $I_{log}(\lambda^2)$ and the renormalization scale dependence expressed by $\ln(\lambda^2)$ cancel out as they should since the model is finite. The arbitrary regularization dependent finite parameter $\alpha$ appeared as the Euclidian version of (\ref{RC})in the calculation of $\delta \Pi_{\mu \nu}$ and should be understood as a free parameter of the model. 

\section{Symmetry Constraints}
\label{Ward Identities}
 As discussed in \cite{Jackiw1}, in theories where radiative corrections are finite, arbitrary parameters stemming from cancellation of divergent integrals should be fixed by symmetries of the underlying model or phenomenology. This is exactly the situation we are confronted with in the calculation of the Coulomb correction for the graphene a.c. conductivity. Spatial $O(2)$ rotational symmetry being preserved in both noninteracting and Coulomb interacting models leads to the following general structure for the polarization tensor \cite{Vafek},\cite{Juricic1}:
 \be
 \Pi_{\mu \nu} (q) = \Pi_A(q_\mu) A_{\mu \nu} + \Pi_B(q_\mu) B_{\mu \nu},
 \ee
with
\be
B_{\mu \nu} = \delta_{\mu i} \Big( \delta_{ij} - \frac{q_iq_j}{\vec{q}^2} \Big) \delta_{j \nu} \,\, {\mbox{and}}
\ee
\be
A_{\mu \nu} = g_{\mu \nu} - \frac{q_\mu q_\nu}{\vec{q}^2} - B_{\mu \nu},
\ee
$\Pi_B (q_\mu)$ being its spatially transverse component.  Hence $\Pi_{\mu \nu} (q)$ is  transverse $q_\mu \Pi_{\mu \nu} (q) = \Pi_{\mu \nu} (q) q_\nu = 0$. Notice that $\Pi_{\mu \nu}^{0}(q)$ in equation (\ref{Pimunu resultado sem beta}) is clearly transverse. Let us study the transversality properties of its $O(e^2)$ correction $\delta \Pi_{\mu \nu}(q)$. For the sake of convenience and comparison with the literature we define \be
\Sigma_{\Omega,p,q} \equiv \int_{\vec{k},\omega}V_{\vec{k}-\vec{p}}G_{\omega + \Omega, k+q},
\label{Sigmapq definicao}
\end{equation}
$\Sigma_{\Omega,p,0}$ being usually called electron self energy in analogy to QED. Using this definition we can formally write
\bq
	&&q^\mu\delta\Pi_{\mu\nu}(q_\mu)= 4\int_{\vec{k},\omega} {\textnormal{Tr}}[G_{\omega, k}\sigma_\nu G_{\omega, k}\Sigma_{p,0,0}] \nonumber \\
	&& - {\textnormal{Tr}}[G_{\omega +\Omega, k+q}\sigma_\nu G_{\omega+\Omega, k+q}\Sigma_{\Omega, p,q}] ,
\label{contracao qmuPcmunu}
\eq
where we have used the straightforward identity
\begin{align}
	q^\mu\sigma_\mu &= -i\Omega {\bf{1}}_{2 \times 2}+\vec{q}\cdot\vec{\sigma} \nonumber \\
	&= G^{-1}_{\omega+\Omega, k+q}-G^{-1}_{\omega, k}.
\label{Contracao qsigma}
\end{align}
Because $\int_\omega (G_{\omega + \Omega, k+q})^2 = 0$ and using the cyclic property of the trace, the study of the transversality of $\delta\Pi_{\mu\nu}$ amounts to investigating the commutator $[G_{\omega+\Omega, k+q},\Sigma_{\Omega, p,q}]$. For this purpose let us further develop the expression for $\Sigma_{\Omega, p,q}$. From (\ref{Sigmapq definicao}), after a Feynman parametrization \cite{Ramond1} for completing the square in the integration variable in the denominator and some lengthy yet direct algebra we can write
\be
\Sigma_{p,q} = \Sigma_{p+q, 0} + e^2 \pi \alpha \vec{\sigma} \cdot \vec{q},
\label{SPQ}
\ee
$\alpha$ being the same arbitrariness that appeared in the computation of $\cal{C}$ given by the (Euclidian version of)  (\ref{RC}). Furthermore following the rules  described before we can demonstrate that $\Sigma_{p+q, 0} \propto \vec{\sigma} \cdot (\vec{p}+\vec{q})$, namely
\bq
\frac{\Sigma_{p+q,0}}{e^2} &=& \frac{\vec{\sigma} \cdot (\vec{p}+\vec{q})}{8} \Big[4 \pi I_{log} (\lambda^2) - \ln \Big( \frac{(\vec{p}+\vec{q})^2}{\lambda^2}\Big) \nonumber \\ &+& 4 \ln 2 - 4 \pi \alpha\Big].
\label{SPPQ}
\eq
We are now in position to discuss the transversality of $\delta \Pi_{\mu \nu}$. Because of (\ref{SPPQ}) it is evident that
\be
[G_{\omega+\Omega, k+q},\Sigma_{p+q,0}] = 0.
\ee
Therefore in view of (\ref{SPQ})
\be
[G_{\omega+\Omega, k+q}, \Sigma_{p,q}] = e^2 \pi \alpha [G_{\omega+\Omega, k+q}, \vec{\sigma} \cdot \vec{q}].
\label{TRANSV}
\ee
As the commutator on the RHS of the equation above does not vanish in general, we are led to conclude that transversality of the Coulomb correction of the polarization tensor implies $\alpha=0$ which in turn fixes ${\cal{C}} = \frac{19 -6\pi}{12}$. Notice that this is the same argument we illustrated for $QED_2$ where gauge invariance fixes the value of the arbitrary parameter to vanish as well.

\section{Discussion}

We proceed to discuss some results for ${\cal{C}}$ that appeared in the literature.
It is straightforward to demonstrate that $\alpha$ evaluates to zero should we employ dimensional regularization to evaluate (\ref{RC}) which in principle indicates that dimensional regularization also leads to ${\cal{C}} = \frac{19 -6\pi}{12}$ contrarily to the result of \cite{Juricic1}.
Direct evaluation of $(\ref{RC})$ in sharp cutoff yields $1/(4 \pi)$ which in view of (\ref{sigmac resultado}) yields  ${\cal{C}} = \frac{25 -6\pi}{12}$ but strikingly violates transversality as the commutator (\ref{TRANSV}) does not vanish. 

Finally it was pointed out in \cite{Juricic1} that a Ward-Takahashi identity similar to the one that appears in field theoretical quantum electrodynamics could be derived defining the ``vertex function" $\Lambda_\mu$ derived from the Fourier transform of the four point matrix function $\pi_\mu (\vec{r_1}-\vec{r},t_1-t,\vec{r}-\vec{r_2},t-t_2) = \langle T_t j_\mu(t,\vec{r})\psi(t_1,\vec{r_1})\psi^\dagger(t_2,\vec{r_2}) \rangle$ such that
\be
\pi^\mu_{\omega,k;\omega+\Omega,k+q} = G_{\omega, k} \Lambda^\mu_{\omega,k;\omega+\Omega,k+q} G_{\omega+\Omega, k+q}
\ee
leading to
\be
q_\mu \Lambda^\mu_{\omega,k;\omega+\Omega,k+q} = \Sigma_{\omega+\Omega,k+q,0} - \Sigma_{\omega,k,0}.
\label{SWI}
\ee
To derive (\ref{SWI}) it is important to notice that the continuity equation was used in its naive form $ \nabla\cdot\vec{j}+\frac{\partial\rho}{\partial t}=0 $ which, according to \cite{Vignale} is no longer satisfied in an interacting system. Assume that (\ref{SWI}) is valid for the Coulomb vertex function to the first order in the coupling constant,
\be
\delta \Lambda_\mu (\Omega,p, q) = - \int_{\vec{k},\omega} \frac{2 \pi e^2}{|\vec{p}-\vec{k}|} G_{\omega, k} \sigma_\mu G_{\omega+\Omega, k+q}
\ee
which according to (\ref{SWI}) should obey
\be
q^\mu \delta \Lambda_\mu (\Omega,p, q) = \Sigma_{\omega+\Omega,p+q,0} - \Sigma_{\omega,p,0}.
\label{SWIC}
\ee
Let us explicitly verify under which requirements this identity is satisfied evaluating separately both sides of (\ref{SWIC}) using the IR framework . After some lengthy yet perspicuous algebra we get, for the RHS
\bq
&&\Sigma_{p+q,0}-\Sigma_{p,0}=e^2\vec{\sigma}\cdot\vec{p}\left\{-\frac{1}{8}\ln{\left[\frac{(\vec{p}+\vec{q})^2}{p^2}\right]}\right\}
\nonumber \\ &&+e^2\vec{\sigma}\cdot\vec{q}\left\{\frac{\pi}{2}I_{log}(\lambda^2)+\frac{1}{2}\ln{2}
-\frac{1}{8}\ln{ \frac{(\vec{p}+\vec{q})^2}{\lambda^2}}\right\} \nonumber \\
&& +e^2\vec{\sigma}\cdot\vec{q}\left(\frac{\pi\alpha}{2}\right).
\label{Lado direito da identidade de Ward}
\eq
whereas the LHS of (\ref{SWIC}) yields
\bq
&& q^\mu \delta \Lambda_\mu =e^2\vec{\sigma}\cdot\vec{p}\left\{-\frac{1}{8}\ln{\left[\frac{(\vec{p}+\vec{q})^2}{p^2}\right]}\right\}
\nonumber \\ && +i e^2 \Omega \Big( \frac{1}{8}- \frac{\pi \alpha}{2}\Big)\nonumber \\ &&+e^2\vec{\sigma}\cdot\vec{q}\left\{\frac{\pi}{2}I_{log}(\lambda^2)+\frac{1}{2}\ln{2}
-\frac{1}{8}\ln{ \frac{(\vec{p}+\vec{q})^2}{\lambda^2}}\right\} \nonumber \\
&& +e^2\vec{\sigma}\cdot\vec{q}\left(\frac{1}{8}\right).
\label{Lado esquerdo da identidade de Ward}
\eq
where $\alpha$ is exactly the same ST that appeared both in the computation of the ${\cal{C}}$ and in the study of the transversality of $\delta \Pi_{\mu \nu}$. Clearly the Ward-Takahashi identity (\ref{SWIC}) is only fulfilled if $\alpha = 1/(4 \pi)$. The latter however is incompatible with transversality of  $\delta \Pi_{\mu \nu}$ which requires $\alpha=0$. 
Curiously enough the value $\alpha = 1/(4 \pi)$ which satisfies (\ref{SWIC}) leads to
${\cal{C}} = \frac{25 -6\pi}{12}$.

\section{Conclusions}

 We employ a regularization independent framework which manifestly preserves arbitrary regularization dependent parameters which should be fixed by symmetries of the underlying model. It is therefore especially tailored  to handle the graphene conductivity calculation which involves contributions that are separately divergent but whose sum is finite and regularization dependent. Based on spatial $O(2)$ symmetry only which is translated into transversality of the polarization tensor, a free parameter $\alpha$ is fixed to zero. This unambiguously results
in ${\cal{C}}= \frac{19-6\pi}{12} \approx 0.01$ which is in agreement with experimental findings so far \cite{Science} and agrees with dimensional regularization because $\alpha$ evaluates to zero if explicitly evaluated within this regularization (see also \cite{Fogler}). On the other hand, if evaluated with a sharp cutoff, for instance, $\alpha = 1/(4 \pi)$ which breaks transversality and leads to ${\cal{C}}= \frac{25-6\pi}{12} \approx 0.51$. It is important to mention that no recourse to the vertex Ward-Takahashi identity whose validity in the presence of interactions has been contested \cite{Vignale} was made:  transversality is enough to fix the only free parameter of the model.

\acknowledgments{
 Marcos Sampaio and M. C. Nemes thank CNPq-Brazil for financial support. G. Gazzola and A. L. Cherchiglia acknowledge a fellowship from FAPEMIG-MG-Brazil.}


\end{document}